# Light-Emitting Electrochemical Cells of Single Crystal Hybrid Halide Perovskite with Vertically Aligned Carbon Nanotubes Contacts

*Pavao Andričević[1], Xavier Mettan[1], Márton Kollár[1], Bálint Náfrádi[1], Andrzej Sienkiewicz[1,2], Tonko Garma[3], Lidia Rossi[1], László Forró\*[1], Endre Horváth\*[1]*

1Laboratory of Physics of Complex Matter (LPMC), Ecole Polytechnique Fédérale de Lausanne, Centre Est, Station 3, CH-1015 Lausanne, Switzerland;

2ADSresonances SARL, Route de Genève 60B, CH-1028 Préverenges, Switzerland;

3Power Engineering Department, Faculty of Electrical Engineering, Mechanical Engineering and Naval Architecture, University of Split, Split, Croatia

ABSTRACT

Based on the reported ion migration under electric field in hybrid lead halide perovskites we have developed a bright, light emitting electrochemical cell with $CH_3NH_3PbBr_3$ single crystals directly grown on vertically aligned carbon nanotube (VACNT) forests as contact electrodes. Under the applied electric field, charged ions in the crystal drift and accumulate in the vicinity of the electrodes, resulting in an *in operando* formed p-i-n heterojunction. The decreased interface energy barrier and the strong charge injection due to the CNT tip enhanced electric field, result in a bright green light emission up to 1800 cd/m$^2$ at room temperature (average ≈ 60 cd/m$^2$). Beyond the light emission, this original device architecture points to the possibility of implementing vertically aligned CNTs as electrodes in operationally-stable perovskite-based optoelectronic devices.

INTRODUCTION

In the last decade semiconducting organic-inorganic lead halide perovskites, such as methylammonium lead triiodide, $CH_3NH_3PbI_3$ (MAPbI$_3$), and methylammonium lead tribromide $CH_3NH_3PbBr_3$ (MAPbBr$_3$), seem to become one of the most potent game-changers in the photovoltaic industry.[1-4] Most recently, apart from their use in next generation solar cells,[5] fast photon detection, even at low illumination intensities,[6] gas sensing,[7] promising thermoelectric figure of merit,[8,9] memristive effects,[10] optically-switched ferromagnetic behavior in magnetic ions doped MAPbI$_3$ single crystals has also been demonstrated.[11]

For solar cells the detailed balance equation requires external luminescence efficiency close to 100% to attain the Shockley–Queisser limit of ≈33.5%, thus implying that an excellent solar cell material must also be an excellent light emitter. As a result, perovskites have rapidly transitioned from breakthrough materials for solar cells to exceptional semiconducting materials with wide-range of applications in light emission.[12] High photoluminescence combined with compositional flexibility place perovskites as robust technological candidates distinguished by their high color purity (FWHM ≈ 20 nm),[12] bandgap tunability to cover the entire spectrum of visible light[13,14] and low-to-moderate ionization energy (IE) to form stable functional interfaces.[15]

Therefore, lead halide perovskites were successfully integrated into light-emitting optoelectronic prototype devices. Tunable amplified spontaneous emission and lasers,[16,17] light-emitting field effect transistors,[18] and light emitting diodes (LEDs) from the infrared[19] to bright-green range have been demonstrated.[20] A typical *state-of-the-art* perovskite LEDs possess a complex, multilayer device architecture, which usually consist of: a hole-transport layer, a light-emitting sheet of 3D layered[19,20] or nanostructured[21,22] perovskite, an electron-transport layer and electrical contacts. The use of electron- and hole-transport layers is considered to be crucial for perovskite LEDs because it lowers the electron/hole injection energy barriers, resulting in low operation voltage and high electroluminescence (EL) efficiency.[23] However, fabrication of such devices is more complicated, necessitates utilization of orthogonal solvents and air-sensitive transport layers. Finding alternative material for electrodes is therefore essential for commercial applications of perovskite based optoelectronic devices.

Here we present a fairly simple architecture for light emitting device, which consists only of two components, methylammonium lead tribromide (hereafter $MAPbBr_3$) and vertically aligned carbon nanotube (VACNT) forests as contact electrodes. The operational principle is based on ion migration under electric field. The space charge at the contacts reduces the barrier for charge injection and the electrons and holes recombine in the crystal resulting in light emission. We take advantage of the strong electric field enhancement at the nanotubes tip for increased charge injection (up to 3 mA) into the crystal. The basic principle of our device is very similar to the mechanism of light-emitting electrochemical cells (LECs).[24-29] Light emitting devices employing a broad range of active layers with ion conductivity from polymer to recently perovskite/ionic electrolyte composites. However, this simple single crystal device presents an unprecedented color purity for an LEC (full width at half maximum of 7 ± 1 nm at 20 K) and the operation extends to room temperature achieving an average brightness of ≈ 60 cd/m² (50 µA) up to flashes of maximum brightness of 1800 cd/m² at 2.8 mA. Knowing the ease for growing both MAPbBr$_3$ and VACNT, this result represents a viable route for future light emitting devices, as well expanding the family of hybrid perovskite LECs.

RESULTS AND DISCUSSION

Perovskite light emitting devices were fabricated by immersing VACNT forests into a saturated solution of MAPbBr$_3$. Under the inverse temperature crystal growth conditions,[30] the fast-growing single crystal gradually protruded and engulfed the individual carbon nanotubes (CNTs). Elemental analysis by EDX in Figure S1 shows clearly the interfaces and the engulfed, overlapping region between CNTs and MAPbBr$_3$, resulting in a 3-dimensionally enlarged MAPbBr$_3$/VACNT junction.[31] Such process, repeated on the opposite side of the perovskite single crystal, allow vertically aligned CNTs to latch also onto this facet, thus forming a symmetric device architecture, with double VACNT electrodes (Figure 1a and b). This is our central device (most of the discussion is addressed to it), but for the sake of comparison, an asymmetric contact device was also prepared with VACNT and one silver electrodes.

An important feature of our device is field emission from VACNT, since carbon nanotubes are well-known to act as field emitters in vacuum[32] or in a dielectric, in this case MAPbBr$_3$. Field emission is a charge-injection mechanism occurring when electron tunnel through a potential barrier under a bias lower than the breakdown voltage. Experimentally, its signature is observed in the I-V measurements, as provided by the empirical Fowler and Nordheim equation:[33] $I = AV^2 \exp(-B/V)$, where $A$ and $B$ depend on the geometry and local environment of the field-emitters, and on the dimensions of the device. The linear relationship between $\ln(I/V^2)$ and $1/V$ in the inset of Figure 1c and in Figure S2 thus supports field emission from VACNTs into MAPbBr$_3$, at low- and ambient- temperatures. Interestingly, the slope (the coefficient *B*) in the inset varies with the sweep direction, pointing towards a change of the geometry of the tunneling barrier.

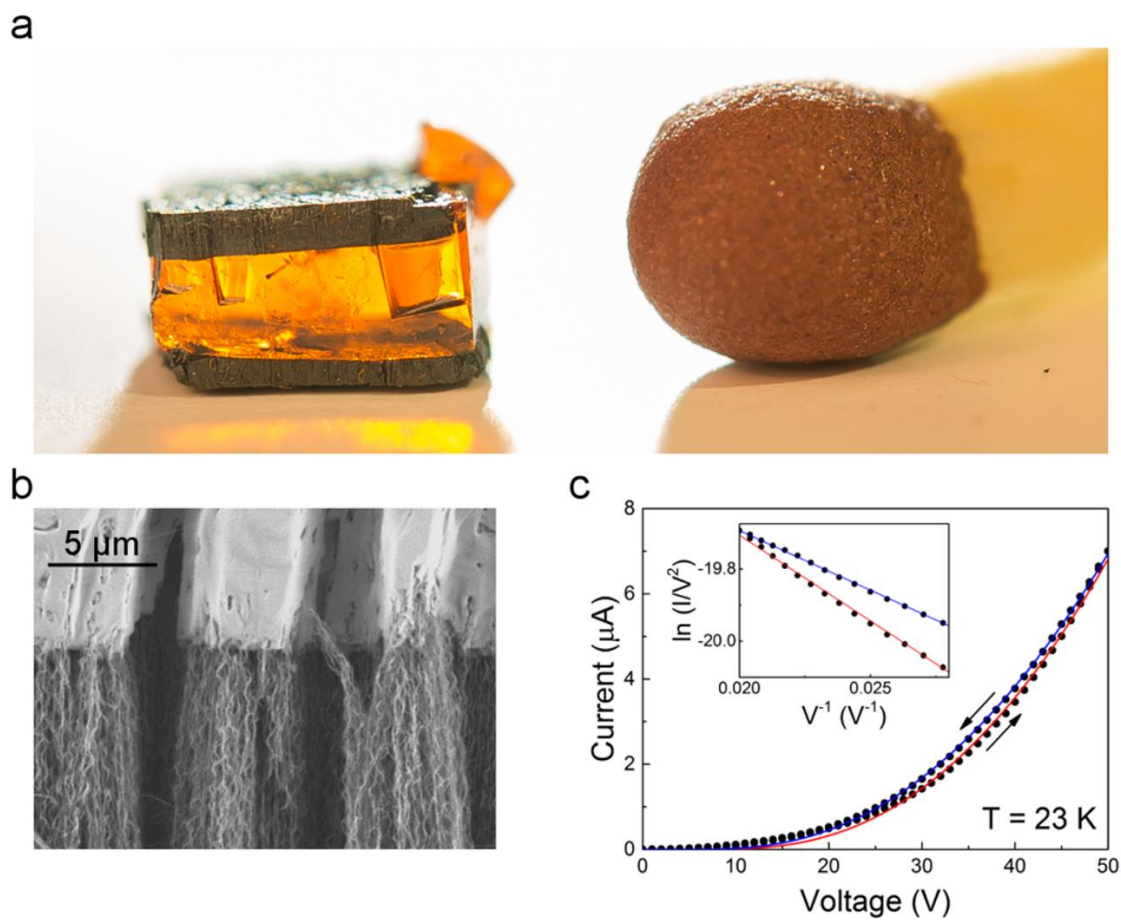

**Figure 1.** Presentation of the Light-Emitting Electrochemical Cell. (a) Optical image of a MAPbBr$_3$ single crystal sandwiched between two VACNT electrodes, resulting in a symmetric architecture (the match gives the scale); (b) SEM image of the interface between MAPbBr$_3$ and the engulfed VACNT electrodes; (c) Current-voltage characteristics of a symmetric device. The inset shows an essential feature of the device: the very efficient charge injection from CNTs by field emission represented by the Fowler-Nordheim plot.

The I-V characteristics for both the symmetric and asymmetric device architectures were collected under white light illumination. In our previous work, it was shown that these devices (asymmetric photodiodes) could serve as very sensitive light detectors, capable of detecting low light intensities down to nano-Watts.[31] One could notice, that hysteresis is present in the I-V characteristics and has been associated with the inertia of ion migration in forward and reversed biases.

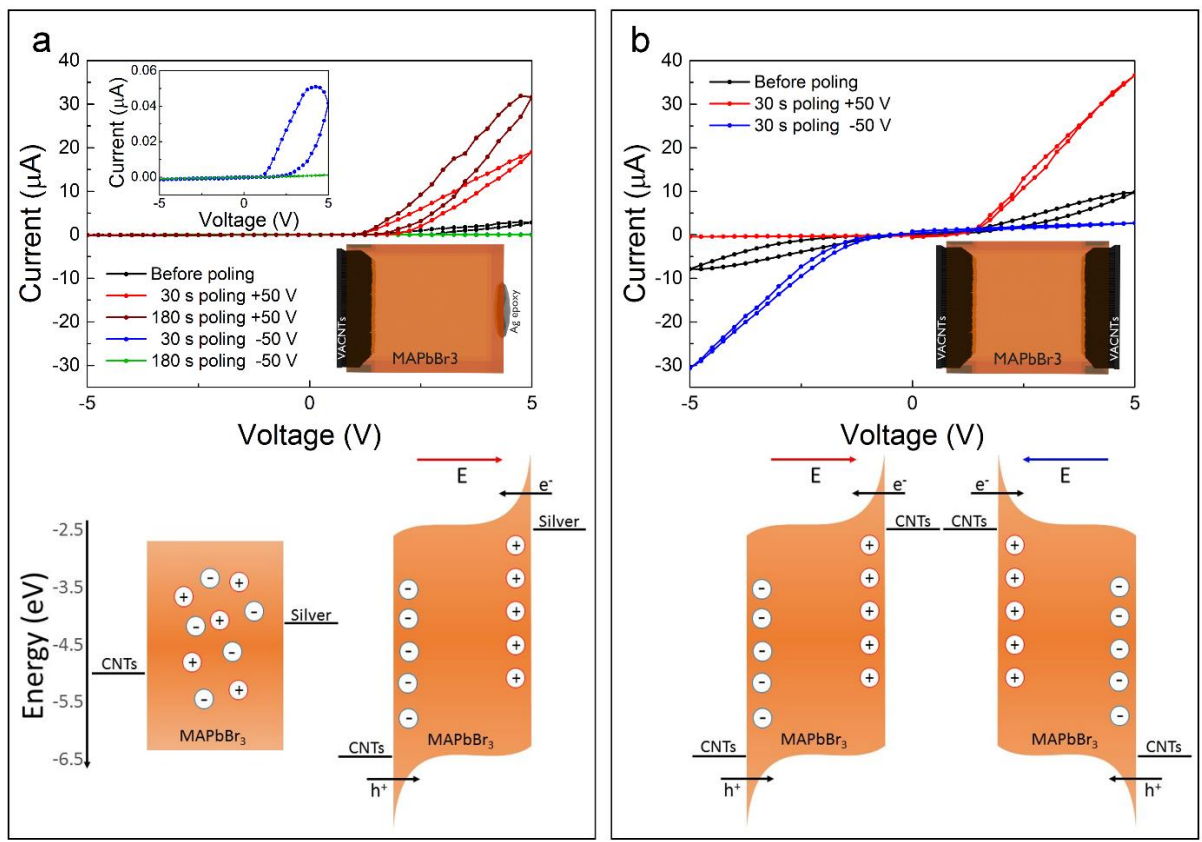

**Figure 2.** The I-V characteristics of the (a) asymmetric device (sketched in the inset) and (b) of the symmetric device (sketched in the inset). The schematic representation of the energy level of the contacts and the position of the ions in open circuit conditions and under an applied external electric field is given below the I-V curves. They were collected under visible light illumination (intensity 1.02 mWcm$^{-2}$), before and after poling treatment, with applying different bias voltage polarities. The voltage was swept from 0 V to +/- 5 V and back. The current intensities through the device

attain lower or higher values, while increasing or decreasing bias voltage, respectively. Inset (a): Suppression of diode characteristics of the asymmetric device after reverse bias poling.

The characterization is performed by I-V measurements for both types of devices before and after performing voltage pre-conditioning, called poling, creating space charge regions.[10,34] As shown in Figure 2 a and b , the I-V characteristics acquired before poling treatment (black) were different for the two types of devices because of the differences in the work functions ($\Phi$) of silver and CNTs. Silver has a $\Phi$ of 4.2 eV,[35] whereas that of CNTs is 5.05 eV.[36] Therefore, it is expected that the asymmetric Ag/MAPbBr$_3$/VACNT device (Figure 2a) has diode like characteristics, allowing current to flow in the forward bias direction while suppressing the flow at reversed polarity due to the high Schottky barrier. After a constant bias voltage is applied for a short period of 30 s, the I-V scans performed at rate of 2 V s$^{-1}$, are different from those before poling. When the device is poled under forward bias, the diode characteristics are preserved with an increase of current from 2 µA to 32 µA at 5V (red). Interestingly, when a constant bias voltage of opposite polarity is applied, the diode characteristics are suppressed (blue) and even nullified (green) in the applied field range (from 2 µA falls to 2 nA at 5V), obtaining symmetric I-V characteristics after 180 s of poling (inset Figure 2a).

The modification of the I-V characteristics is even more pronounced for the device with two VACNT electrodes (Figure 2b). Having symmetrical junctions, it has Schottky I-V characteristics for both positive and negative bias voltages. However, if a constant bias voltage is applied for 30 s just before a quick voltage sweep (2 V s$^{-1}$), the I-V characteristics completely change to a typical diode-like behavior. Now, current flows for voltages of the polarity of the previously applied bias. These changes as well as the time of relaxation back to the initial states depend on the time of voltage pre-conditioning and on the value of the bias voltage. The effects of poling in detail, as well as the changes of I-V characteristic under different times of poling (10 seconds to 10 minutes) and poling voltages (1 to 200 V) are shown in Figure S3. Moreover, it is possible to change the polarity of the open circuit voltage as seen in Figure S3e and f, achieving a switchable photovoltaic effect.[10]

In open-circuit conditions, ions of opposite signs are uniformly distributed within the active layer (Figure 2a). After a constant bias voltage is applied, the ions drift towards their respective electrodes and they accumulate at the perovskite/metal interface forming p-i-n heterojunction structures, increasing the injection rate of electrons and holes from the electrodes to the perovskite layer and governing the devices I-V characteristics (Figure 2b). It is worth to mention that hysteresis was observed in perovskite solar cells, as well, attributing it to ion migration inside the perovskite layer.[37-40]

At bias voltages ($V_b$) larger than 5 V, the current through the device increased over three orders of magnitude (from nA to µA) before saturating. Remarkably, at ambient conditions and starting from the current intensities through the device of ~50 µA, a bright green EL was first observed for the asymmetric device (Figure S4a). Voltages much larger than the band gap were used to achieve this critical current for light emission. Nevertheless, due to the large thickness of our single crystal devices (>1 mm), the external electric field was never larger than 200 V mm$^{-1}$.

To achieve light emission for both polarities, the focus had been moved to the symmetrical device architecture employing VACNTs as both electrodes on the perovskite single crystal. The device, previously conditioned to show diode like behavior, exhibit light emission with further poling at room temperature in ambient conditions. For this device architecture, light emission is visible for both polarities as shown on Figure S4b. Interestingly, the depletion region, which arises due to the high electric field (~10 V mm$^{-1}$), which is the region where the bright light emission is observed, is always near the interface with the lower potential, as was the case for the asymmetric device.

Bright green light is emitted at room temperature in the form of repetitive flashes of 100-120 Hz. In 20 seconds of acquiring light intensity and simultaneously measuring the current through the device, more than 40 excitations can be visible from multiple locations near the active surface of the device (Figure 3a). This yields an average luminance of ~60 cd/m$^2$ for current as low as 50 µA. At higher currents, more charges are injected into the single crystal, resulting in brighter (max. 1800 cd/m$^2$ for 2.8 mA current) and more frequent flashes (Figure 3b). Importantly, stable and continuous light emission has not been observed at ambient conditions. It is likely that the space charge created by the migrated ions in electric field is exposed to thermal fluctuations. The fluctuations change the barrier structure in time, resulting in a fluctuating injected current and light intensity. Current instability at a constant voltage has been common for MWCNT field emitters.[32] However, these emitters observe switching between discrete current levels at low currents, reaching more stable emission at higher currents. Whereas, the perovskite light emitting device exhibits random fluctuations increasing with the value of current. Nevertheless, this is still an open question and should be taken into further consideration.

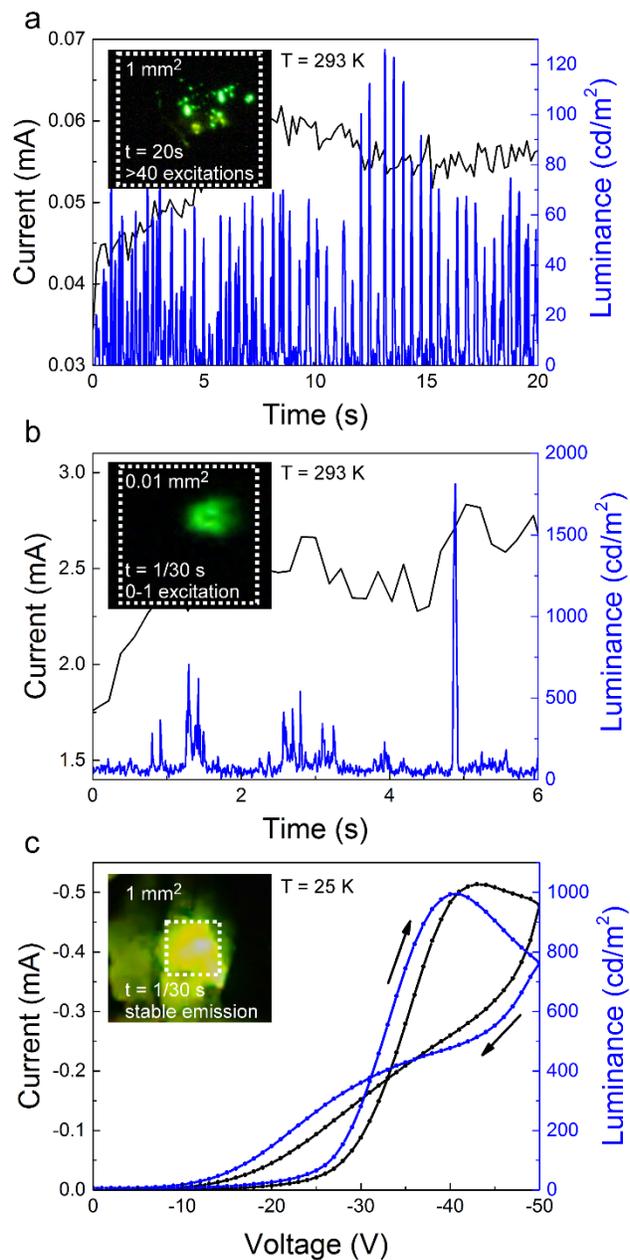

**Figure 3.** (a) Luminance-current in time for an applied bias voltage of 20 V at room temperature. Inset: Optical image of > 40 flashes of light of light during 20 seconds at room temperature. (b) Luminance-current in time for an applied voltage of 40 V (> 2 mA current) at room temperature. Inset: Screen shot of a video of the light emission at room temperature showing the emitting surface of less than 250 μm$^2$. (c) Luminance-current versus voltage at cryogenic temperatures (25 K). Inset: Screen shot of a video of the light emission showing stable light emission with an active emitting surface of 1 mm$^2$.

These fluctuations are damped by cooling the system down to cryogenic temperatures, because the mobility of ions is strongly reduced. The intensity of the injected current became stable, producing a continuous light emission. For example, at 25 K a very bright light emission was observed as seen in the inset of Figure 3c. Luminance-current versus applied voltage (LIV) of the symmetric LEC device for both negative and positive voltages is shown in Figure 3c and S5, respectively. Since the light emission comes from the recombination of electrons and holes in the perovskite, it is expected that the EL intensity correlates with the injected current in $MAPbBr_3$. This is fully confirmed by the LIV measurements at 25 K as well as their dependence with temperature (Figure S5). Continuous light emission stopped at 100 K, but flashes of light with quick changes of intensities could be followed up to room temperature. The device showed stable light emission for up to one hour operation (Figure S5f). The light emission at room temperature, in the form of flashes, and at low temperatures, as a constant light in the dependence of voltage, can be seen in the video in the supporting information.

The spectral analysis of the emitted light at 20 K was performed for turn-on voltages in the range of 100 to 200 V and -30 to -40 V. Already by naked eyes, one can see different colors of the emitted light for the two cases (see inset to Figure 4). Two sets of spectra with similar intensity count numbers were chosen of both polarities and fitted with a Gaussian function (Figure 4). One can read from these spectra two important information: i) the redshift by 20 nm of the negative bias spectra compared to the spectra acquired for positive voltages; and ii) the very narrow full width at half maximum (FWHM), of the order of 8 nm, which represents a high spectral purity. This latter fact is certainly due to the high crystallinity of the sample. In polycrystalline sample or in polymer LECs the spectra are broader multiple times due to the inherent disorder.[41] When applying 150 V, bright green spots with an emission peak centered at 542 nm wavelengths have the energy of the band gap. The -40 V biased case gives a red-shifted signal at 562 nm. It is likely that the different colors are due to different locations of recombination of the electron-hole pairs in the crystal. Our conjecture is that the green light is coming mostly from the crystal surface, while the yellowish from the bulk. Deep in the crystal, after recombination, the light is absorbed and re-emitted in cascades before reaching the spectrometer, and during this process the slight energy loss results in a redshift of the spectra. On the other hand, for positive voltages the recombination takes place near the surface, the light is directly emitted towards the spectrometer at a wavelength of 542 nm.[30] Similar arguments were put forward for red-shifted spectra in photoluminescence measurements by Wenger et al.[42] All the acquired EL spectra with their associated currents are given in Figure S6.

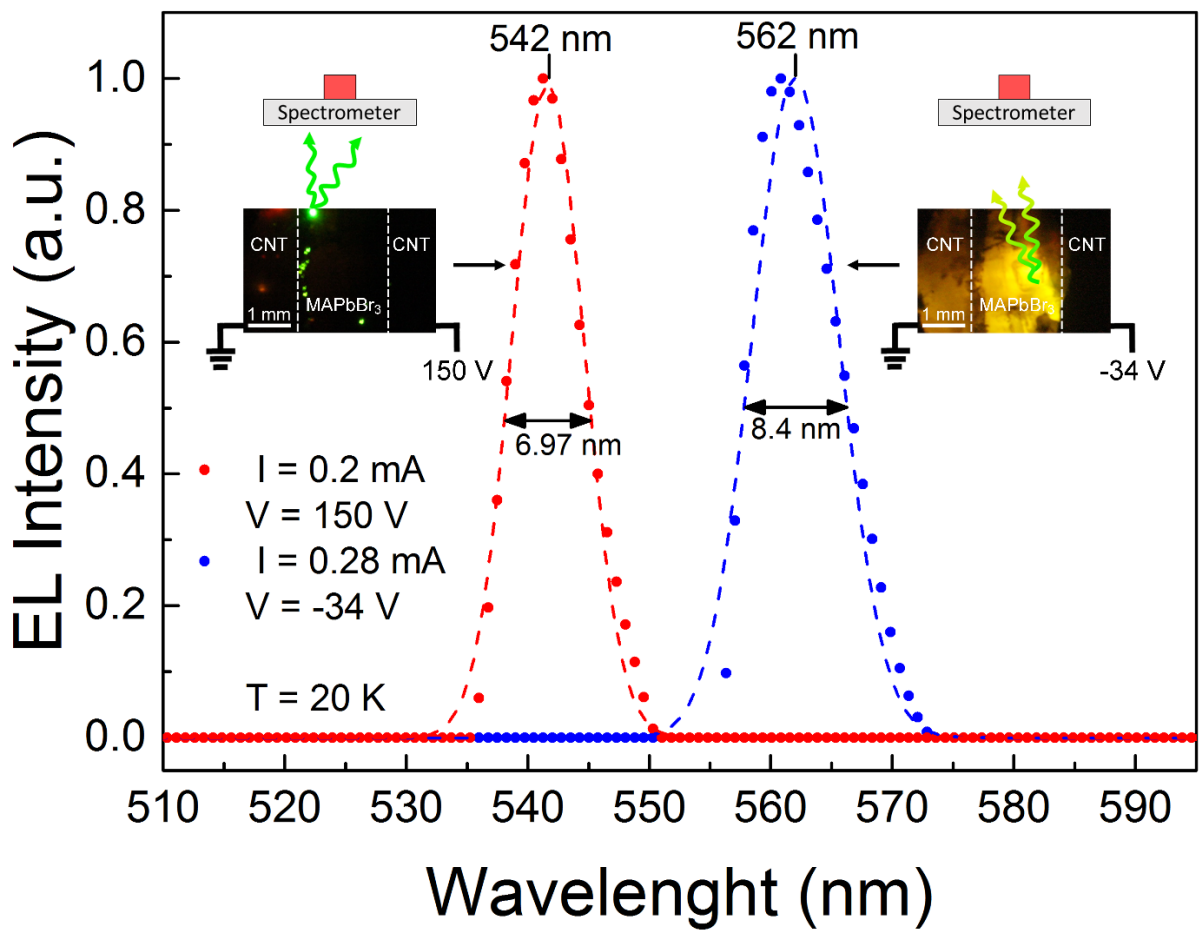

**Figure 4.** The typical normalized EL spectra recorded at 20 K in vacuum, at positive (red, +150 V) and negative (blue, -34 V) bias voltages, fitted with Gaussian function (dashed lines). Inset: Schematic illustration of the measurement setup with the photos of the light emission for the associated EL spectra.

In order to obtain a more complete picture, about the limiting factors of the EL phenomenon in the symmetric architecture, light emission were collected with increasing temperatures from 23 K to 40 K for the -40 V biased case (Figure 5a), and spectral analysis was performed (Figure 5b). At 23 K two EL emission peaks were detected: one at 562 nm with a FWHM of about 7 nm, which is the same as in Figure 4, and a second broad peak at 590 nm with a FWHM of 10 nm. With increasing temperature, the intensity of the peak located at 562 nm weakens with no significant shift in wavelength and FWHM, while the peak centered at 590 nm increases in intensity, simultaneously undergoing a redshift and broadening, achieving a FWHM of 30 nm at 40 K (Figure 5c). We suspect that this peak emerges from radiative recombination at trap states in the perovskite single crystal. The increasing intensity with temperature corroborates with this attribution, as the population of these trap-states grows with temperature. With further increase in temperature, higher absolute values of bias voltages and longer spectrometer integration times were needed to acquire spectra curves. This was possible only for the second peak at 590 nm up to 60 K as shown in Figure S7. The redshift and the increasing FWHM is coming from the interaction of the created photons with lattice vibrations. These features are confirmed on the asymmetric device recorded in the 40 - 300 K range shown in Figures S8 and S9.

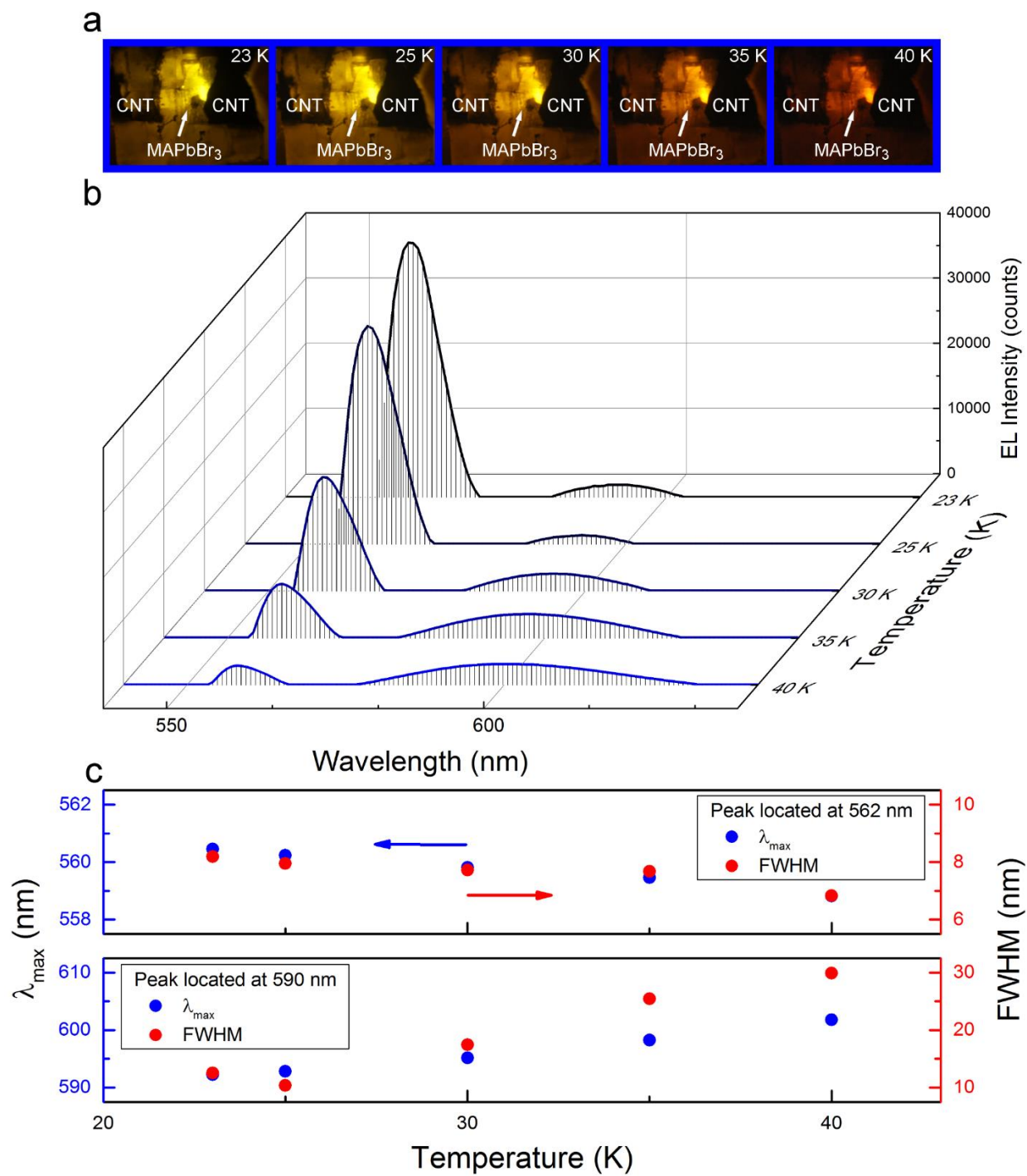

**Figure 5.** (a) Optical images of the light emission with -40 V of bias voltage as a function of temperature, highlighting the redshift. (b) Spectral analysis of the EL spectra reveal an emission peak at 562 nm and at 592 nm at 23 K, which have different temperature dependences. (c) The temperature dependence of the wavelength and FWHM extracted from a Gaussian fit.

In the above documented EL, the poling, that is the movement of ions under electric field is primordial (like in polymeric LECs), the building up of space charge layer near the contacts which facilitate the charge injection into the crystal. To have a better insight into the timescale of this phenomenon, we have studied the time-dependence of the current increase upon applying a constant external voltage of 10 V during 10 hours (Figure S10). The monotonic increase of current was fit by a tri-exponential function giving three time constants. The shortest one (65 s) corresponds to the halogen ion movement, the intermediate (35 min) to the MA+ diffusion, and the longest one (7 h) to the slow thickening of the space charge layer. Such a hierarchy of the diffusion activation energies was found by Puscher et al.[43], as well. It has to be mentioned that no detectable degradation of the device was observed during the 10 working hours of operation at ambient atmosphere.

A further characterization of the device is done by measuring its capacitance ($C$) via the discharge current (similarly to Zhang et al.[44]) after poling of the VACNT/MAPbBr$_3$/VACNT sample (thickness 1 mm, surface area 10 mm$^2$). A certain bias (e.g., 2 V) was applied for a short period of time (≈15 s) in dark to accumulate the ions at the perovskite/CNT interfaces, that is to charge the $C_i$, which represents the areal capacitance from the blocking of ions at the electrode interfaces.[38] The device was then rapidly switched to zero bias to measure the discharge current (Figure S11a, inset). The accumulated charge density is the time integral of the discharge current. From the measured charge densities of the accumulated ions at the interfaces under various biases (Figure S11a), one can extract a value of $C_i$ ≈1 µF cm$^{-2}$. This large areal capacitance exceeds that of conventional electrostatic capacitors[45] with a typical value of 0.1 µF cm$^{-2}$. Furthermore, as shown in Figure S11b, the areal capacitance increases with the poling time. These results suggest that the organic–inorganic perovskites could serve as potential solid-state supercapacitors.

In general, the electroluminescence devices are very popular, using a broad range of active layers which possess the ionic conductivity.[24] Mostly these are conjugated polymers, ionic transition-metal complexes, small-molecules, quantum-dots, to recently used perovskite nanoparticles.[46] The latter one was used by Aygüler and co-workers, using MAPbBr$_3$ and FAPbBr$_3$ perovskites mixed with an ion polyelectrolyte (IP) to increase the ionic conductivity.[47] Following up this contribution, light emission was achieved in mixtures of perovskite nanoparticles with poly(ethylene oxide) (PEO) polymers,[41] composite films[48] or as thin films using ITO/PEDOT:PSS and MoO3/Au electrodes. For the sake of completeness, the best-in-class perovskite-based LECs and LEDs is shown in Table S2.

CONCLUSION

Here we demonstrated the operation of new, simple LECs based on MAPbBr$_3$ and VACNTs. Our VACNT/MAPbBr$_3$/VACNT device is the first room temperature single crystal light emitter using symmetrical metallic electrodes without electrolyte or additional n-or p-type selective layers. Due to the single crystalline nature of the active material, the spectral purity of the emitted light is very high at low temperatures. The oriented carbon nanotubes by field emission inject a high current density (due to the tip enhanced electric field) which gives bright light emission up to 1800 cd/m$^2$ even at room temperature. The here developed design points to the possibility of implementing vertically aligned CNTs as electrodes in operationally-stable optoelectronic devices. The VACNT-based technology can also serve as a versatile platform for future electrode development.

EXPERIMENTAL SECTION

*VACNT growth*: Carbon nanotubes were grown by catalytic chemical vapor deposition. Firstly, transition metals, such as Fe, Co, Ni, or their alloys were used as the catalysts, while $Al_2O_3$, MgO, and $SiO_2$ as oxide supports. VACNT are then grown utilizing a special preparation protocol documented in our previous work[31,49,50] named the supergrowth method. The CNT self-assemble into vertically oriented cellular arrays during the growth on the substrate as high as 2.5 mm. The VACNT forest contain predominantly multi-wall CNTs and are 95% porous allowing the direct grow of perovskite SC on the VACNT forest.

*Crystal growth*: Crystals of the methylammonium lead tribromide were synthesized by solution growth. The lead (II) acetate trihidrate (3.3 mmol, $Pb(ac)_2$ x $3H_2O$, >99.9%) was reacted with saturated HBr solution (6 ml, 48 wt% HBr in $H_2O$). The formed $PbBr_2$ precipitate is stable in the acidic solution. The respective amount (3.30 mmol) of methylamine ($CH_3NH_2$) solution (40 wt% in $H_2O$) was pipetted into the 5 °C ice cooled solution of $PbBr_2$. The cold solution avoids the evaporation of methylamine during the exothermic reaction. Orange colored microcrystallites of $CH_3NH_3PbBr_3$ were formed. The $MAPbBr_3$ crystals were recrystallized in a temperature gradient of 15 °C in the acidic media to get transparent, high purity crystals.

$MAPbBr_3$ single crystals were grown by inverse temperature crystallization from its saturated solution in DMF. $MAPbBr_3$ (0.8 g) was dissolved per $cm^3$ of DMF at room temperature. The substrate was immersed in the solution, and a $MAPbBr_3$ seed crystal was placed on the top of the VACNT. Crystal growth was initialized by increasing the temperature of the solution from room temperature to 40 °C with a heating rate of 5 °C per hour. We observed that the fast-growing seed single crystals gradually protruded and engulfed the individual nanotubes. These types of inclusions, when the original form of the included mineral is preserved in the host crystal is categorized as a protogenetic inclusion in mineralogy. The slow heating rate suppress the formation of new seed crystals in the supersaturated solution and our $MAPbBr_3$ seed crystal can grow rapidly on the substrate. When the required size of crystal was reached, the $MAPbBr_3$ – VACNT composite was removed from the solution, wiped and dried. The same growth method was repeated using the $MAPbBr_3$ – VACNT junction as a seed crystal to obtain two VACNT electrodes on the perovskite single crystal.

*Optoelectronic characterization.* All measurements of the I-V characteristics of the device and dependence to poling at room temperature were done in ambient conditions. The device characteristics have been determined by two-point resistivity measurements, using golden wires contacted with Dupont® 4929 silver epoxy as electrical leads. A Keithley 2400 source meter allowed us to measure the current with < 0.1 nA resolution, while tuning the applied bias voltage, in dark and under visible light illumination.

*Intensity measurements of the light emission at room temperature:* A Thorlabs PM100D Compact Power and Energy Meter Console with its corresponding software were used to obtain the intensities of light with < 0.1 nW resolution. The light emitting device was contacted with two tungsten needles as electrical leads. A Keithley 2400 source meter was use to apply voltage to the sample and measure current with < 0.1 nA resolution. The photodetector was positioned 1 cm from the sample at a small angle to simultaneously capture optical images or videos of the bright green light with a Canon EOS 600D camera (ISO1600). All measurements were done at room temperature and in ambient atmosphere.

*Intensity and spectra measurements of the light emission at low temperatures*: To achieve stable light emission the light device was mounted inside a close cycle cryostat with optical windows and brought to cryogenic temperatures (20 to 80 K) and high vacuum. A Keithley 2400 source meter was again use to apply voltage to the sample and measure current with < 0.1 nA resolution. The dependence of the light intensity with voltage and in time were determined by measuring the photocurrent of a Thorlabs FDS100 silicon photodiode, which was positioned ~5 cm from the sample behind an optical window. Likewise, the spectra of the emitted light were measured using an Ocean Optics QE65 Pro Spectrometer. The light was collected using a series of lenses to an optical fiber, positioned on the cryostat optical windows. The spectra were acquired using the Ocean Optics SpectraSuite software. Care was taken to limit the current (max. 2 mA). With increasing temperatures, the absolute value of the voltage had to be adjusted (increased) in order to keep a sufficient current, required for light emission. Luminance of the device was calculated from the spectra at room and cryogenic temperatures as shown in Figure S12. Due to many losses in our setup (optical windows etc.) much lower values are obtained compared to direct measurements of intensities of the same device at room temperature. Optical images clearly show a much brighter light so the intensities obtained by the silicon photodiode is calibrated as shown in the Supporting Information.

## ASSOCIATED CONTENT

Supporting Information

I-V characteristics of the light emitting device under different pooling schemes as well as room to cryogenic temperature characterization of the asymmetric LEC device.

## AUTHOR INFORMATION

Corresponding Authors

Email: laszlo.forro@epfl.ch

Email: endre.horvath@epfl.ch

ACKNOWLEDGMENT

This work was supported by the Swiss National Science Foundation (No. 513733) and the ERC advanced grant "PICOPROP" (Grant No. 670918). The authors gratefully acknowledge Dr. Daniel Oberli for the technical support in obtaining the light intensity.

For Table of Contents Use Only

Light-Emitting Electrochemical Cells of Single Crystal Hybrid Halide Perovskite with Vertically Aligned Carbon Nanotubes Contacts

Pavao Andričević, Xavier Mettan, Márton Kollár, Bálint Náfrádi, Andrzej Sienkiewicz, Tonko Garma, Lidia Rossi, László Forró*, Endre Horváth*

Electroluminescence was observed of hybrid halide perovskite single crystals at ambient conditions, with symmetric 3-dimensionaly enlarged vertically aligned carbon nanotube electrodes. Light emission was discussed in the framework of the working principles of light emitting electrochemical cells.

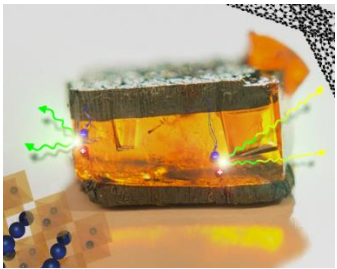